\documentstyle[twocolumn,aps,prl,epsf]{revtex}

\begin{document}

\title{Spin Analogs of Proteins: Scaling of ``Folding'' Properties}

\author{Trinh Xuan Hoang, Nazar Sushko, Mai Suan Li, and Marek Cieplak}

\address{Institute of Physics, Polish Academy of Sciences,
Al. Lotnikow 32/46, 02-668 Warsaw, Poland }
\address{
\centering{
\medskip\em
{}~\\
\begin{minipage}{14cm}
Reaching a ground state of a spin system is analogous to a protein
evolving into its native state.  We study the ``folding'' times for
various random Ising spin systems and determine characteristic
temperatures that relate to the ``folding''.  Under optimal kinetic
conditions, the ``folding'' times scale with the system size as a
power law with a non-universal exponent.  This is similar to what
happens in model proteins.  On the other hand, the scaling behavior of
the characteristic temperatures is different than in model proteins.
Both in the spin systems and in proteins, the folding properties
deteriorate with the system size.
{}~\\
{}~
\end{minipage}
}}

\maketitle
\newpage

\section{Introduction}

Recent numerical studies \cite{Shakhnovich,Cieplak} indicate that
characteristic folding times, $t_{fold}$, of model proteins grow with
the number of aminoacids, $N$, as a power law with an exponent which
is non-universal -- it depends on the class of sequences studied and
on the temperature.  The resulting deterioration of the folding
properties also manifests itself in the way in which temperatures that
relate to folding scale with $N$ \cite{Cieplak}.  There are two such
characteristic temperatures: $T_f$ and $T_{min}$.  The first of these
is a measure of the thermodynamic stability  -- it can be defined
operationally as a temperature at which the probability to occupy the
native (the lowest energy) state crosses $\frac{1}{2}$. The second
temperature is one at which the folding kinetics is the fastest. At
temperatures, $T$, below $T_{min}$, glassy effects set in. Aminoacidic
sequences that correspond to proteins should have a $T_f$ that is
bigger than $T_{min}$, or at least comparable to $T_{min}$.  Otherwise
the sequences are bad folders.  Studies \cite{Cieplak} of two and
three dimensional lattice Go models \cite{Go} of proteins suggest that
$T_{min}$ grows with $N$ whereas $T_f$ first grows and then it either
saturates or it grows at a lower rate than $T_{min}$. There exists
then a characteristic size, $N_c$, at which $T_{min}$ starts exceeding
$T_f$ and for $N > N_c$ the sequences necessarily become bad folders.
This suggests existence of a size related limit to physiological
functionality of proteins.

The question we ask in this paper is to what extent the scaling
behavior of $t_{fold}$, $T_f$, and $T_{min}$ that was found in the
lattice Go model of proteins is typical or, in other words, what are
the classes of universality for these quantities.  Specifically, we
consider Ising spin systems: uniform ferromagnets, disordered
ferromagnets and spin glasses.  Disordered ferromagnets has been shown
recently \cite{Garstecki} to have a phase space structure, as
described by the so called disconnectivity graphs
\cite{Becker,Cambridge,Cambridge1}, quite akin to that characterizing
proteins, at least for a small number of spins, $N$.  Spin glasses, on
the other hand, have been found to have the phase space structured as
in random sequences of aminoacids which are bad folders.  The spin
systems do not ``fold'' but an evolution into their ground states can
be considered to be analogous to the folding process
\cite{Garstecki,Frauen} and $t_{fold}$ can be defined as the
characteristic time needed to pass through the ground state for the
first time, which generally does not coincide with a relaxation time.
Thus $t_{fold}$, $T_f$, and $T_{min}$ can be determined like for the
proteins and we may additionally enquire how do $T_f$ and $T_{min}$
relate to the effective critical temperature as determined from the
specific heat and magnetic susceptibility.

Another motivation to consider the ``folding'' in spin systems is that
the analogies between spin systems and proteins have already permeated
the language in which the physics of proteins is couched.  It is not
clear, however, to what extent these analogies are accurate when it
comes to actual details.  One qualitative concept, in this category,
is that of the energy landscape \cite{landscape,landscape1}: spin
glasses are said to have rugged energy landscapes but proteins should
have a landscape which is much smoother and funnel-like.  Another such
concept is frustration \cite{Bryngelson}: the structural frustration
in proteins should be ``minimal'' whereas the frustration in the
exchange couplings leads to the slow kinetics as found in spin
glasses.  These concepts have been probed, e.g. in the random energy
model \cite{Bryngelson} which again originated in the context of spin
glasses \cite{Derrida}.

The basic message of this paper is that the spin -- protein analogies
are indeed valid but the details of the behavior are usually distinct.
What is analogous, for instance, is that the folding times have a
characteristic U-shaped dependence on $T$ \cite{Socci}. Furthermore,
the folding properties are the best for small system sizes  and then
they deteriorate with $N$. In particular,  the ``folding'' times at
$T_{min}$ in spin systems do grow as a power law with $N$. On the
other hand, both $T_{f}$ and $T_{min}$ of simple spin systems
generally decrease with $N$ and the nature of the phase transition is
not a finite size version of the first order as is the case with the
proteins.

The origins of the difference between spin systems and the Go models
of proteins in the behavior of $T_f$ and $T_{min}$ remain to be
elucidated.  It should be noted that there are no kinematic
constraints on flips of any spin whereas the possible moves in the
protein folding process must preserve the chain connectivity and they
have to depend on the actual conformation and thus on the history. The
constrained character of the protein dynamics makes it acquire aspects
of the packing problem, especially so if the native state is maximally
compact -- such as considered in the studies of scaling in model
proteins.  The packing aspects become insignificant when dealing with
longer and longer $\alpha$-helices \cite{Hoang}. We illustrate this
point here by considering a 2-dimensional lattice version of the
$\alpha$-helices (H) as described within the Go scheme and show that
these objects indeed become behaving like spin system when $N$ becomes
bigger and bigger.  Notice that the helices have the monomer-monomer
interactions of a local kind.  Thus the energy barrier against
unfolding essentially does not depend on $N$ which is not expected of
structures with more complex contacts.

Most of this paper, however, will be focused on systems described by
the Ising spin Hamiltonian: 
\begin{equation}
{\cal H} \; \; = \; \; - \sum_{<ij>} J_{ij} S_i S_j \; \; ,
\end{equation}
where $S_i = \pm 1$, and the exchange couplings, $J_{ij}$, connect
nearest neighbors on the square and cubic lattices with the periodic
boundary conditions. There are $L^D$ spins in the system where $D$
denotes the dimensionality and $L$ the linear size of the system.
We consider four models of the exchange couplings: 
1) spin glasses (SG) in which the $J_{ij}$'s are numbers drawn from the
Gaussian probability distribution with a zero mean and a
unit dispersion; 
2) uniform ferromagnets (FM) with
$J_{ij}=1$; 
3) the disordered ferromagnets (DFM) with $J_{ij}$ chosen
as the absolute values of the Gaussian numbers;  
4) the weakly disordered ferromagnets (DFM')  with the $J_{ij}$'s
being random numbers between 0.9 and 1.1.  We find that it is the
latter system which is the most protein-like.

In Sections 2 and 3 we discuss the $T$ and $N$ -dependencies of the
``folding'' times respectively.  In Section 4 we present results on
the scaling behavior of $T_f$ and $T_{min}$ in systems SG, FM, DFM,
DFM', and H.  Finally, in Section 5, we demonstrate that the
temperatures we study are quite distinct from the critical temperature
of the spin systems.

\section{Temperature dependence of ``folding'' times}

The concepts used in this paper are illustrated in Figure 1 which
shows the temperature dependence of the characteristic ``folding''
time in the $D$=2 Ising spin systems considered.  In each category,
data for a representative example system are shown.  We obtain
$t_{fold}$ by a standard Monte Carlo process in which one typically
starts from 1000 random initial spin configurations and determines the
median time to reach the ground state for the first time.  The spins
are updated sequentially and the ``folding'' times are given in Monte
Carlo steps per spin, i.e. the total number of spin updates
divided by the number of spins. In the SG case, the ground state (or
at least its close approximation) is obtained by multiple slow
annealing processes followed by a quenching procedure.  The general
shape of the $T$-dependence is like in the protein systems -- it
corresponds to a U-shaped curve with a minimum at $T_{min}$. The bigger the
disorder, the higher the $T_{min}$ -- the DFM system has the highest
$T_{min}$ among the systems with the ferromagnetic ground state. 
Figure 1 also shows that the SG system has a lower value of $T_{min}$
than the DFM. However, this does not reflect the degree of the disorder 
since the two systems are different in nature.  
The fact that the SG system has a lower value of $T_{min}$ than the
corresponding DFM system is related to the fact that local energy
barriers against spin flipping are generally higher in a DFM than in a
SG due to a nonzero value of the average exchange coupling.
The phase space structure of the uniform ferromagnet is so simple,
containing few local energy minima, that the low temperature upturn
does not develop down to $T$=0. In this case, we shall attribute a
zero value to $T_{min}$. The similar phenomenon has been observed in
the lattice Go model with repulsive non-native contacts \cite{Li} for
which only few local minima are available.  The shortest ``folding''
time, $t_{min}$ corresponds to $t_{fold}$ that is determined at
$T_{min}$.

\begin{figure}
\epsfxsize=3.2in
\centerline{\epsffile{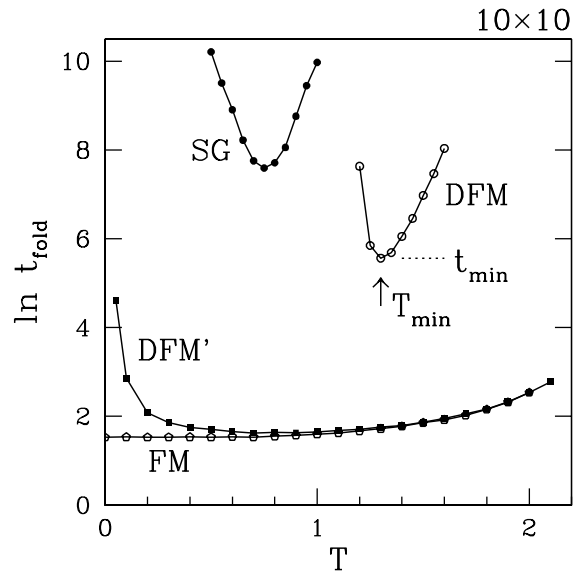}}
\caption{The temperature dependence of the characteristic ``folding''
time for the two dimensional FM, DFM', DFM and SG Ising systems.
Here, $L$=10 and the number of starting configurations is equal to
1000.}
\end{figure}

The temperature dependence of the folding time on $N$ is also computed
for the Go model of ``helices'' on the two-dimensional square lattice. 
A ``helical'' native state for $N$=16 in shown at the top of Figure 2.
The meanders shown in the figure become longer and longer when $N$
grows.  The Hamiltonian for the system is given by
\begin{equation}
{\cal H}\;=\;\sum _{i<j} \; B_{ij} \; \Delta _{ij} \;\; ,
\end{equation}
where $\Delta _{ij}$ is either 1 or 0 depending on whether the
monomers $i$ and $j$ are nearest neighbors on the lattice but
not nearest neighbors along the chain, or not.  When $\Delta _{ij}$ is
non-zero, the two monomers are said to form a contact.  The definition
of the Go model is that $B_{ij}$ is 1 for the native contacts (such as
seen in Figure 2) and 0 otherwise.  Thus the properties of the system
are determined entirely by the native conformation.  The dynamics are
defined in terms of a Monte Carlo process which satisfies the detailed
balance conditions as explained in \cite{Henkel,Ukraine,Cieplak}.

\begin{figure}
\epsfxsize=3.2in
\centerline{\epsffile{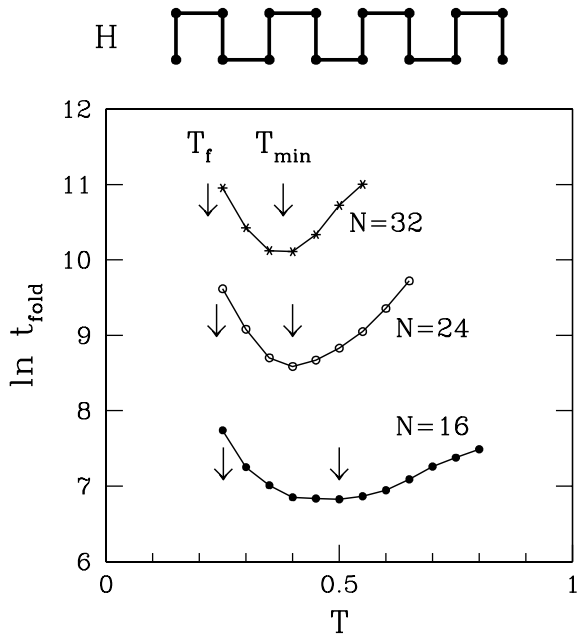}}
\caption{The top of the figure shows an example of a ``helical''
conformation on 2$D$ lattice used in the paper. The conformation shown
is for $N$=16.  The lower part shows the dependence of $t_{fold}$ on
$T$ for three indicated lengths of the ``helix''. The values of $T_f$
and $T_{min}$ are marked by the arrows.}
\end{figure}

The lower part of Figure 2 shows the characteristic U-shape dependence
of $t_{fold}$ on $T$ for system H. What is different compared to the
models of maximally compact proteins is that the positions of both
$T_{min}$ and $T_f$ are seen to go down with $N$ -- the point to which
we shall come back in Section 4.

\section{Scaling properties of folding times at $T_{min}$}

In order to study the scaling properties of disordered systems, such
as the spin systems with random exchange couplings, one needs to
consider ensembles of samples with properties which are similar
statistically. Thus for each $N$ we have considered up to 50 samples
and for each we have performed simulations of ``folding'' in the Monte
Carlo process. The median folding times as a function of $T$ have been
calculated for each sample separately and we determine their fastest
folding condition.  Typically it is done by considering 1000 folding
trajectories at each $T$. But for the FM and DFM' systems with a small
size up to 40000 trajectories at each $T$ have been used due to the
broadness of the minimum.  The value of $t_{min}$ has been determined
at $T$=$T_{min}$ that corresponded to a given sample and only then the
average of $t_{min}$ over samples has been calculated.  Figure 3 shows
the scaling of the average $t_{min}$ for the 2$D$ systems: FM, DFM',
DFM, SG and H.  Figure 4, on the other hand, deals with the 3$D$ Ising
systems.

\begin{figure}
\epsfxsize=3.2in
\centerline{\epsffile{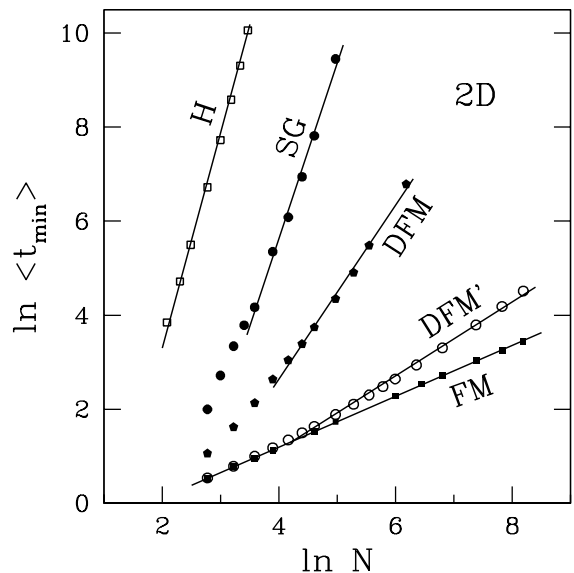}}
\caption{The scaling of $t_{min}$ for 2$D$ FM, DFM', DFM,
SG and the ``helices''.}
\end{figure}

All of the results are consistent with the power law:
\begin{equation}
\left<t_{min}\right> \sim N^{\lambda} \;\;\; ,
\end{equation}
where the values of $\lambda$ are shown in Table 1.  Interestingly,
$\lambda$ for the spin systems depends much more strongly on the type
of the spin system than on its dimensionality. On the other hand,
in the Go models of proteins with the maximally compact native state,
the dependence on $D$ is strong: it is of order 6 and 3 in 2 and 3 $D$ 
respectively \cite{Shakhnovich,Cieplak}. 

It should also be noted that for the 2$D$ lattice ``helices'', 
$\lambda \approx 4.59 $ is substantially smaller than the exponent
found for the Go proteins with the maximally compact native state
which points to the role of the packing effects.

\begin{figure}
\epsfxsize=3.2in
\centerline{\epsffile{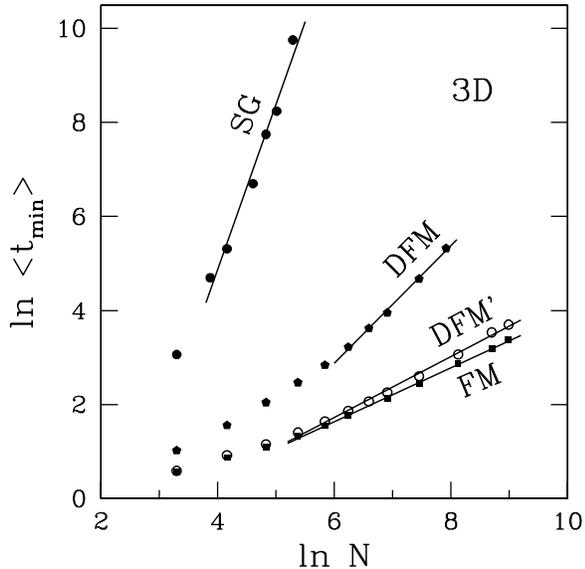}}
\caption{Scaling of $t_{min}$ for 3$D$ FM, DFM', DFM and SG.}
\end{figure}

\begin{figure}
\epsfxsize=3.2in
\centerline{\epsffile{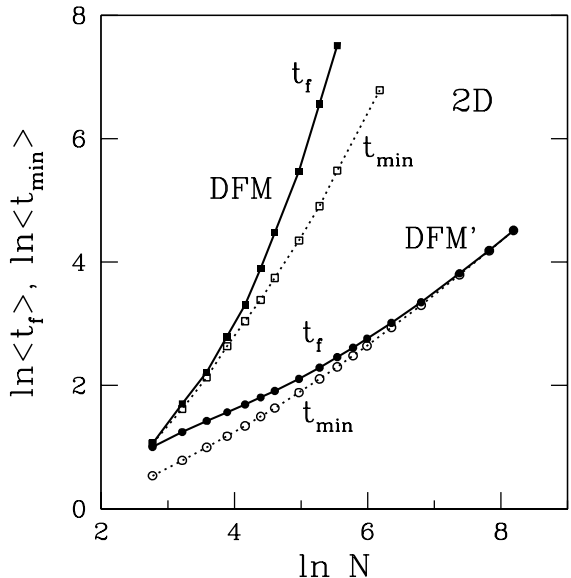}}
\caption{Scaling of $t_{f}$ for 2$D$ DFM and DFM' (solid lines)
compared to the scaling of $t_{min}$ (dotted lines). The effective
slope for the last three data points corresponding to DFM is 3.54.}
\end{figure}

The strong dependence of $\lambda$ on the choice of the exchange
couplings is similar to the lack of universality found in model
proteins \cite{Shakhnovich}. Also in analogy to the models of
proteins, the scaling exponent depends on the temperature.  Figure 5
shows that $t_{fold}$ evaluated not at $T_{min}$ but at $T_f$ grows
with an even bigger exponent or possibly the growth becomes
exponential.  This emphasizes the optimality of the kinetics at
$T_{min}$.  In the DFM' case $t_f$ and $t_{min}$ merge together
because, as we shall see in the next section,  the temperatures $T_f$
and $T_{min}$ merge themselves.

\vbox{
\begin{table}[t]
\begin{center}
\begin{tabular}{c c c}
System  &   $\lambda$ ($D$=2)   &  $\lambda$ ($D$=3) \\
\hline
FM   &  $0.54 \pm 0.01$  &  $0.57 \pm 0.01$ \\
DFM' &  $0.79 \pm 0.01$  &  $0.65 \pm 0.02$ \\
DFM  &  $1.85 \pm 0.05$  &  $1.25 \pm 0.05$ \\
SG   &  $3.73 \pm 0.15$  &  $3.52 \pm 0.25$ \\
H    &  $4.59 \pm 0.08$  &  -             \\
Go   &  $6.3 \pm 0.2$    &  $3.1 \pm 0.1$\\
\end{tabular}
\end{center}
\caption{The exponent $\lambda$ for the 2 and 3 $D$ spin systems and
for the 2$D$ Go models of helices.  The symbol ``Go'' in the table
refers to the Go models with the maximally compact native states
as studied in Ref. [2].}
\end{table}
}

The possibility of a power law scaling for the folding time has been 
proposed theoretically by Thirumalai\cite{Thirumalai}
(see also Ref. \cite{Camacho}) based on scaling concepts from 
polymer physics combined with some phenomenological assumptions.  
In particular, the power law scaling is argued to be relevant to 
proteins which fold through direct pathways with a nucleation mechanism.
For indirect pathways, the folding time is determined primarily
by activation process with barriers which were argued to
scale as $N^{1/2}$.      
There has been also a number of other studies of how a typical
free energy barrier, $B$, in model proteins scales with the number of
monomers.  All of these studies are phenomenological in nature and the
barrier $B$ is often calculated at the folding transition temperature
$T_f$. One assumes that the folding time is related to the barrier
$B$ through an Arrhenius-like law: $\tau\;\sim\;exp(B/k_BT)$, as it
is typically written for the relaxation time.  
In the random energy
model\cite{Bryngelson}, and also in another mean field approach for
the Go model with a nonspecific critical folding nucleus\cite{Takada}
the barrier scales linearly with $N$.  
Recently, Finkelstein and Badredtinov \cite{Finkelstein},
and also Wolynes \cite{Wolynes2} have proposed a $N^{2/3}$ law by
using a capillarity approximation.  
Gutin's et al. and our power laws
for $t_{fold}$ obtained in simulations of the lattice proteins would
formally correspond to a logarithmic dependence of the barrier on $N$
at the temperature of the fastest folding.  

It should be noted, however, that the physics of folding coincides
with that of equilibration only in the limit of low temperatures
\cite{Henkel}. At high temperatures, for instance, the relaxation
times are short but the folding times are long since the search for
the ground state takes place primarily in the regions of phase space
which are energetically remote from the target native state.  Thus the
behavior of the barriers may have little bearing on the folding times
at $T_{min}$ which corresponds to the crossover between the physics of
folding through equilibration and the physics of folding through a
search for a state that takes place in equilibrium.  At low
temperatures, the roughness of the energy landscape becomes more and
more significant, and the changed nature of the local barriers against
the reconfiguration is expected to affect the scaling laws.
Understanding of the scaling behavior of the folding time at $T_{min}$
and at low temperatures still needs to be worked out -- both in the
protein and spin systems.  The latter systems may prove to be easier
conceptually and computationally.

\section{Scaling properties of $T_f$ and $T_{min}$}

We now discuss the scaling of characteristic temperatures. $T_{min}$
is determined from the kinetic data. $T_f$, on the other hand, is
calculated by starting from the ground state and performing a long run
that determines the equilibrium probability of the system staying in
the ground state.  The probabilities are determined as a function of
$T$ and $T_f$ is obtained by an interpolation to where the value of
1/2 is crossed.  For the spin systems, our results are based on up to
200 ``unfolding'' trajectories which last for  up to 10000 Monte Carlo
steps per spin.

\begin{figure}
\epsfxsize=3.2in
\centerline{\epsffile{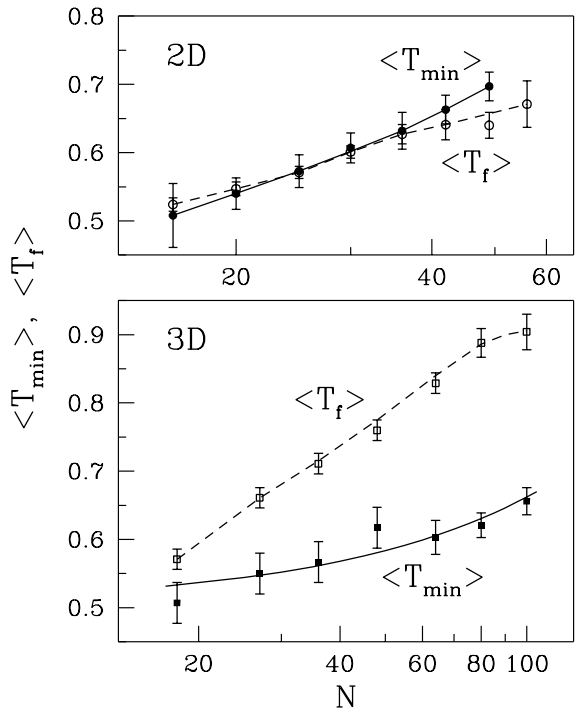}}
\caption{Scaling of $T_f$ and $T_{min}$ for the 2 and 3 $D$ lattice Go
models of proteins. Data points are taken from Ref. [2].}
\end{figure}

As a point of reference, we first consider the scaling properties of
$T_f$ and $T_{min}$ which were found in the two and three dimensional
lattice Go models of proteins \cite{Cieplak}.  The corresponding data
points are now shown, in Figure 6, as a function of $N$ on the
logarithmic scale.  Both $T_f$ and  $T_{min}$ grow with $N$. The data
points suggest that $T_{min}$ grows indefinitely -- the larger the
system size, the higher $T$ is needed to secure the optimal folding
conditions.  On the other hand, $T_f$ appears to tend to a saturation
value -- there is a limit to the thermodynamical stability.  This
finding is consistent with an analytical result obtained by Takada and
Wolynes \cite{Takada} for Go-like proteins studied within a droplet
approximation.

Figure 7 shows the scaling of $T_f$ and $T_{min}$ for the 2$D$ lattice
``helix'' system. At $N\le 8$ the foldability is good but on
increasing the $N$, the behavior is entirely different: the glassy
effects  decrease in importance -- $T_{min}$ goes down -- but also the
thermodynamic stability becomes more and more insignificant.  The
slopes for the $N$-dependence of the two temperatures are somewhat
different and the corresponding plots may cross at some large value of
$N$.  Thus it is possible that good foldability can reappear at some
large values of $N$ -- but at a very low $T$.

\begin{figure}
\epsfxsize=3.2in
\centerline{\epsffile{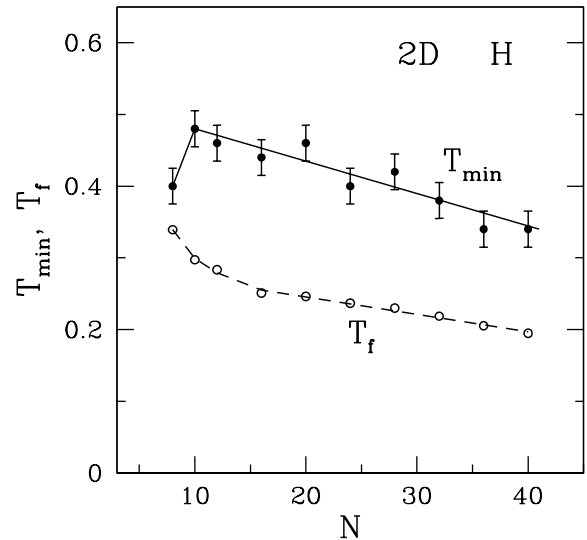}}
\caption{The same as in Fig. 6 but for 2 $D$ Go models of the ``helices''.}
\end{figure}

We now ask what kind of the scaling  behavior of $T_f$ and $T_{min}$
characterizes the spin systems? Figures 8-11, for systems FM, DFM',
DFM and SG respectively, demonstrate that in no case the scaling is
like for the Go lattice models with the maximally compact native state
but in some cases it is akin to the behavior exhibited by the 2$D$
``helix''.

Both for the ``helix'' and for all of the spin systems studied here,
$T_f$ decreases with $N$ monotonically which is not what happens in
the Go models of proteins.  This difference in behavior can be traced
to the following observation.  $T_f$ is defined through the equation
\begin{eqnarray}
P_N \; = \;
\frac{1}{1+ \sum_l'
\exp(-(E_l-E_N)/k_BT_f)} \; = \; \frac{1}{2} \; \; ,
\end{eqnarray}
where $P_N$ is a probability of being in the ground state, $E_N$ is
the energy of the ground state, $E_l$ is the energy of an $l$'th
state, and the sum written in the denominator excludes the ground
state.  At temperatures which do not exceed $T_f$, the sum is
dominated by the low energy excitations. In the spin systems and in
the ``helix'', energies of these excitations do not depend on $N$.
For instance, in the Ising case they are of order $2zJ$, where $z$ is
the coordination number and $J$ denotes a characteristic value of the
exchange interactions. It is only the number of terms in the sum
itself that grows with $N$. This leads to $T_f$ decreasing with $N$.
On the other hand, in the model proteins, the energies of the
excitations typically do depend on $N$ which may have a competing
effect on $T_f$ relative to the impact of the number of states.

We now turn to discussion of the scaling properties of $T_{min}$. From
Figures 8-11 it is clear that it has opposite tendencies for proteins
and spin systems.  For the 2$D$ FM and DFM' systems one observes an
increase followed by a saturation. In all other spin systems, instead
of the saturation, one observes a maximum followed by and asymptotic
decrease.  The difference may reflect the presence of the kinematic
constraints on possible moves in proteins due to their polymeric
nature.  Such constraints may cause emergence of barriers which depend
on $N$ and result in a growing $T_{min}$.  In spin systems such
kinematic constraints do not exist, each spin configuration has $N$
possible ways to move out with a cost which does not depend on $N$.
Such a high number of degrees of freedom gives the spin systems a
large flexibility to cross from local minima to local minima.
Thus there is no potential for an indefinite growth of $T_{min}$.  The
initial growth, in the random systems, reflects on the role of the
growing number of the local energy minima which may form kinetic traps
and make the kinetics glassy. And yet the asymptotic decrease of
$T_{min}$ suggests that the relevant traps do not need a $N$-dependent
energy to overcome or points to some entropic effect.

The case of the ``helix'' system may appear puzzling at a first glance
since it possesses polymeric constraints and yet they do not lead to a
growing $T_{min}$.  Note that in contrast to the model proteins with
maximally compact native states \cite{Cieplak}, the Hamiltonian for
the ``helix'' contains terms related only to the local contacts. Thus
the chain is much more flexible 
it is not tightly packed in its native state.  The energy barriers
against escaping from the traps do not depend on the chain length, and
therefore the ``helix'' exhibits the spin-like properties. High
energetic barriers in proteins are often associated with breaking of
some tertiary contacts.

\begin{figure}
\epsfxsize=3.2in
\centerline{\epsffile{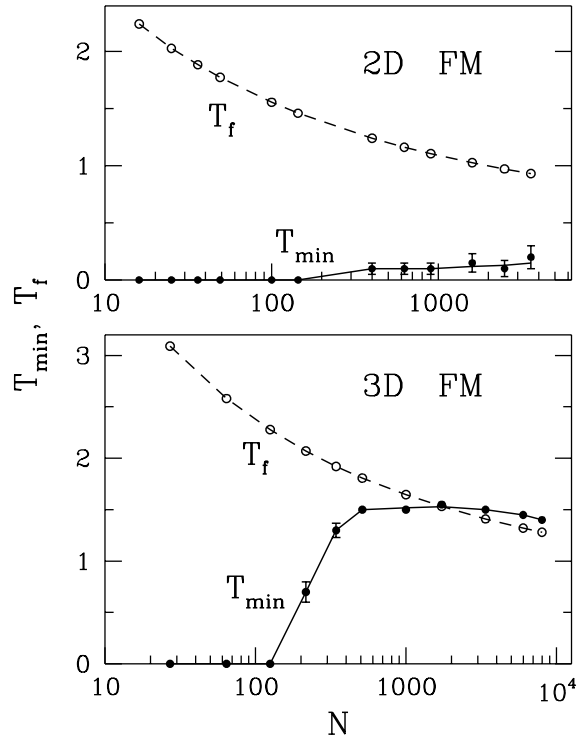}}
\caption{The dependence of $T_f$ and $T_{min}$ of 2 and 3 $D$ FM on $N$.}
\end{figure}

\begin{figure}
\epsfxsize=3.2in
\centerline{\epsffile{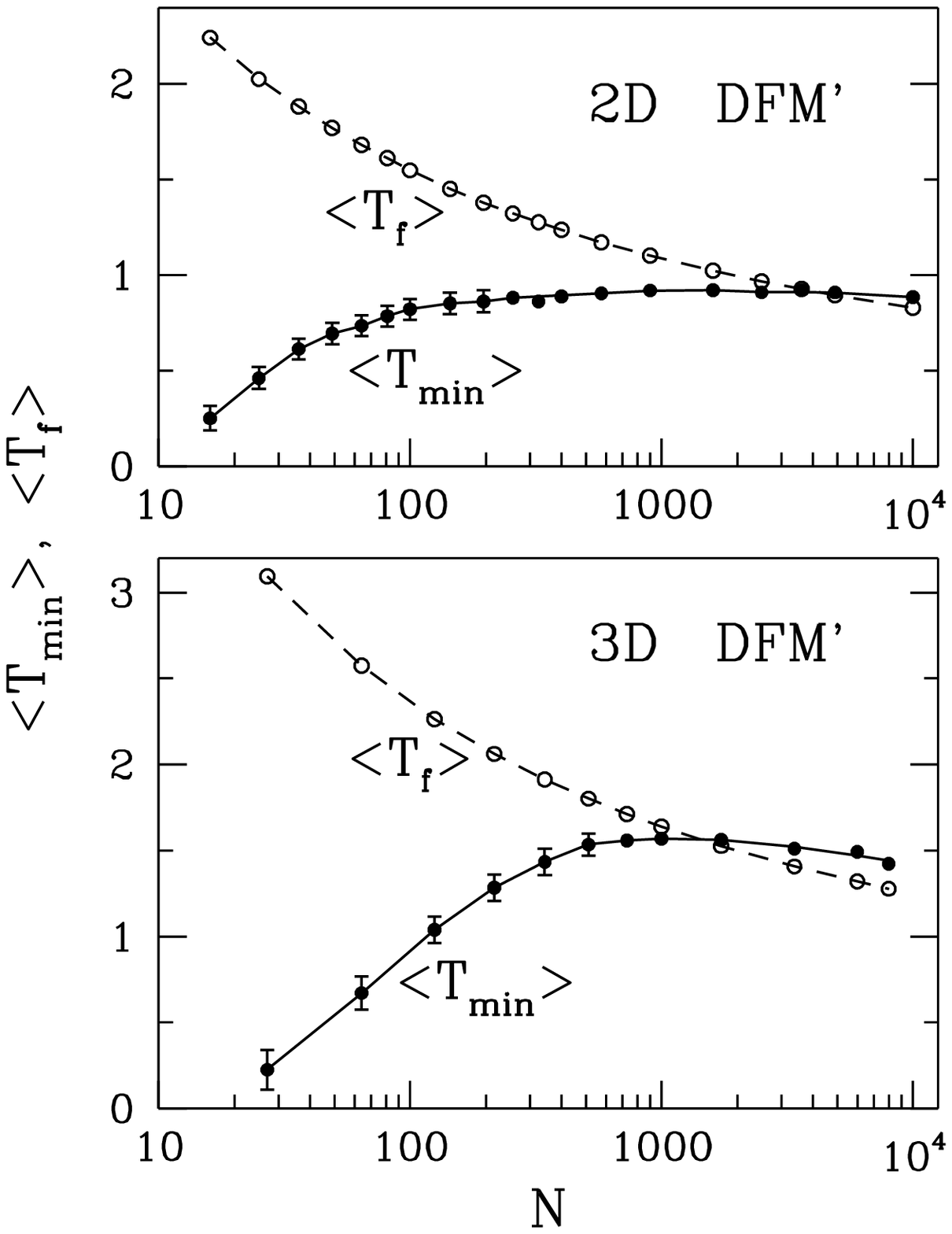}}
\caption{The same as in Figure 8 but for the DFM' systems.}
\end{figure}

\begin{figure}
\epsfxsize=3.2in
\centerline{\epsffile{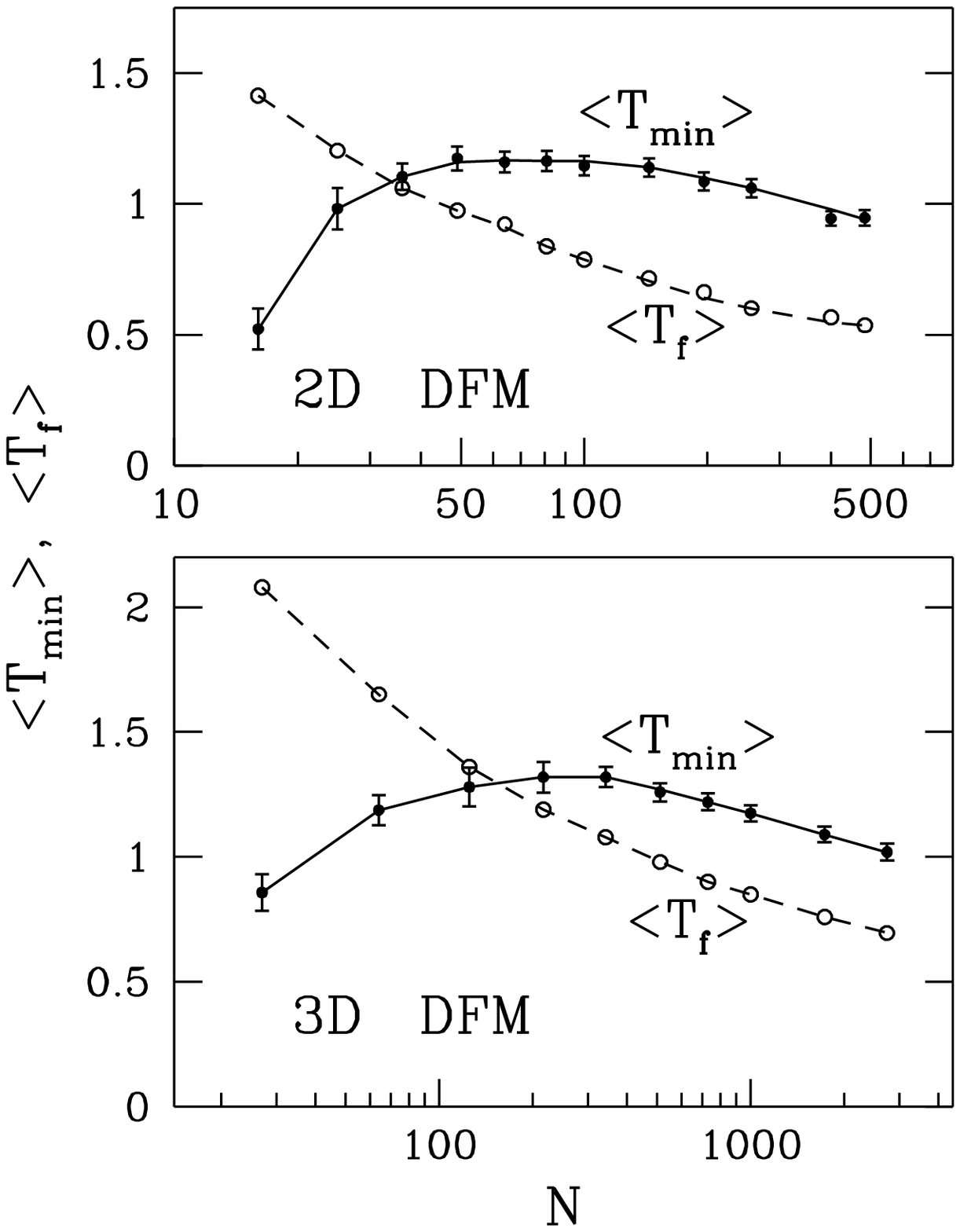}}
\caption{The same as in Figure 8 but for the DFM systems.}
\end{figure}

\begin{figure}
\epsfxsize=3.2in
\centerline{\epsffile{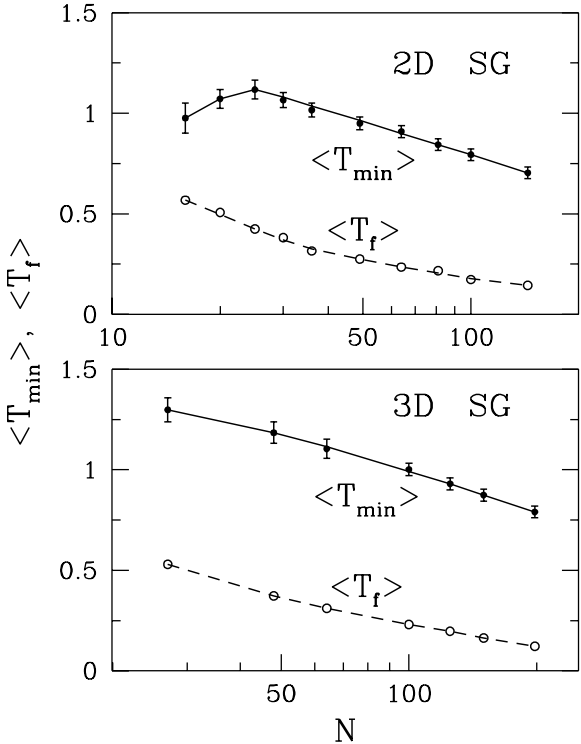}}
\caption{The same as in Figure 8 but for the SG systems.}
\end{figure}

\section{The specific heat and susceptibility}

We now compare $T_f$ and $T_{min}$ to the usual critical temperatures
that characterize spin systems. We focus on the properties of the
specific heat, $C$, and susceptibility, $\chi$, defined through

\begin{equation}
C \; = \; \frac{\left<E^2\right> - 
\left<E\right>^2}{NT^2} 
\end{equation}
and
\begin{equation}
\chi = \; \frac{\left<M^2\right> - \left<M\right>^2 }{NT}
\end{equation}
respectively, where $M$ is the magnetization.
The temperatures at which $C$ and $\chi$ have a maximum
will be denoted as $T_C$ and $T_\chi$, respectively.

In addition, and in analogy to the proteins \cite{Klimov}, we study
the structural  fluctuations
\begin{equation}
\Delta \chi_s = \left<\chi_s^2\right> - \left<\chi_s\right>^2
\end{equation}
which are defined in terms of the structural overlap function 
\begin{equation}
\chi_s = \frac{1}{N} \left|\sum_{i=1}^N S_i \; S^{(N)}_i \right|,
\end{equation}
where $\{S^{(N)}_i\}$ is the spin configuration in the ground state.
(For the ferromagnets $\chi_s$  is the same as the absolute value of
the magnetization per spin).  These fluctuations also have a maximum
at some temperature which will be denoted by $T_s$.  It has been
suggested \cite{Klimov} that for  proteins $T_s$ should be about $T_f$
and a small difference between $T_s$ and $T_C$ is a signature of fast
folding.

All of these thermodynamic quantities are averaged  over 10 to 20
samples and 100 trajectories for each.
In each trajectory, the first $5\,000$ to $10\,000$ Monte Carlo steps
per spin are spent for equilibration. The trajectories were then
further evolved between $20\,000$ and $50\,000$ steps per spin.  The
lower values above refer to the DFM' system and the higher -- to the
SG system.

Figure 12 shows the scaling behavior of $T_C$, $T_{\chi}$, and $T_s$
for the 2$D$ DFM' and SG systems. In the case of DFM', the three
temperatures converge to one common critical temperature. Note that
none of these temperatures has anything to do with $T_f$ or $T_{min}$.
In the SG system, $T_s$ and $T_{\chi}$ tend to separate asymptotics
than $T_C$ but again none of these temperatures coincides with $T_f$
or $T_{min}$.  The physics of folding is not related to the critical
phenomena.  It should be noted that a phase transition is spin glasses
shows as a singularity in the nonlinear susceptibility. 
In the 2$D$ SG system, the peak position in the nonlinear
susceptibility should be at $T$=0 for any system size\cite{Rieger}.

\begin{figure}
\epsfxsize=3.2in
\centerline{\epsffile{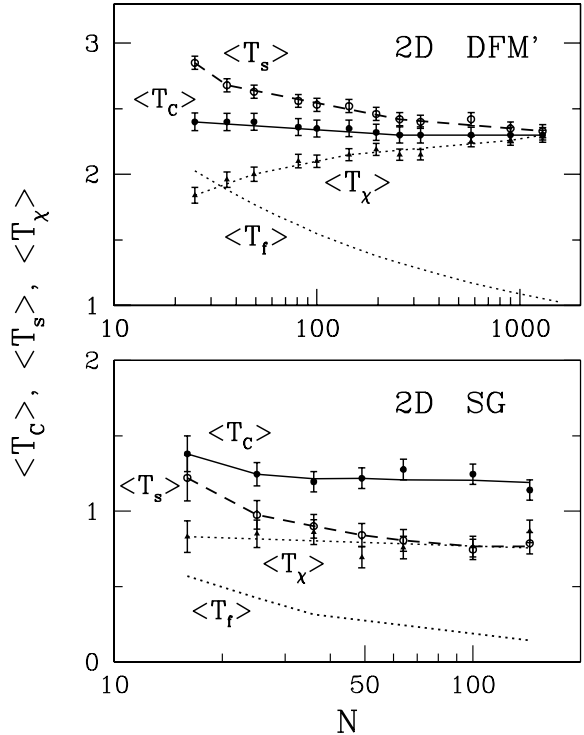}}
\caption{The size dependence of $T_C$, $T_{\chi}$, $T_s$, and $T_f$
for the 2$D$ DFM' and SG systems.}
\end{figure}

In summary, we have studied Ising spin systems from the perspective
of protein folding. We have demonstrated that there exist many 
similarities between the spin and polymeric systems. In particular,
we have shown that both kind of systems have the property
of a power law scaling of the folding time at $T_{min}$ as a 
function of $N$. We point out that this holds independent
of whether the system is a good or bad folder and is thus some
universal feature of folding.

Among the random spin systems studied here, the DFM' systems have the
biggest range of the small $N$ values at which $T_f$ is larger than
$T_{min}$, both in 2 and 3 $D$. Thus these small sized systems are
the best analogs of good folders and can serve as toy models that
mimic the physics of proteins.  Spin glasses of any size, on the other
hand, do indeed mimic the physics of random heteropolymers.
Asymptotically though, each random spin system is a bad ``folder''.

This work was supported by KBN (Grant No. 2P03B-025-13).
M.S.L. thanks H. Rieger for useful discussions.


\end{document}